# Polarization-Differential Loss Enabled High Polarization Extinction in Hollow-Core Fibers

Yizhi Sun[1,2,3], Shoufei Gao[1,2,3,†], Xiangqi Wang[1,2], Xianhao Qi[1,2], Zhixi Liang[1,2], Rui He[1,2], Wei Ding[1,2,3,4], Yingying Wang[1,2,3,*]

[1]Guangdong Provincial Key Laboratory of Optical Fibre Sensing and Communication, Institute of Photonics Technology, Jinan University, Guangzhou 510632, China

[2]College of Physics & Optoelectronic Engineering, Jinan University, Guangzhou 510632, China

[3]Linfiber Technology (Nantong) Co., Ltd. Jiangsu 226010, China

[4]Pengcheng Laboratory, Shenzhen 518055, China

Email: †gaosf@jnu.edu.cn; *wangyingying@jnu.edu.cn

**Abstract:** Delivering a well-defined state of polarization over hollow-core fibres (HCFs) is pivotal for next-generation ultra-stable photonic systems. Yet in all existing HCFs, whether birefringent or not, their polarization extinction ratio (PER) rapidly deteriorates during propagation or under mechanical disturbance, leaving no practical high and stable PER solution. Here, we break this impasse by embedding a polarization-differential loss (PDL) mechanism directly into the cladding architecture. By deliberately phase-matching one core polarization to a leaky cladding resonance, the unwanted polarization is strongly attenuated while the orthogonal mode remains low-loss, providing passive stabilization that complements birefringence. A 19 μm-core prototype sustains a PER of 30 dB across a 45 nm band after 725 m, with just 8.5 dB km$^{-1}$ loss on the working axis and a peak PDL of 456 dB km$^{-1}$. Reducing the core to 15 μm boosts the PDL to ~1800 dB km$^{-1}$ and delivers PER ≈ 40 dB over 102 m, exceeding 50 dB within a 5 nm window while remaining stable when coiled to 6 cm radius. To our knowledge, this is the first demonstration of sustainably high PER in near-kilometer-scale HCFs under practical perturbations. PDL-engineered HCFs thus overcome the longstanding polarization bottleneck, opening a scalable path to bend-tolerant, single-polarization fiber links for gyroscopes, coherent laser delivery, and timing networks.

## 1. Introduction

Polarization is a fundamental attribute relevant to the transverse wave nature of light and underscores a plethora of modern photonics applications, enabling advanced functionalities ranging from high-efficiency data encoding, precision measurement of physical quantities to interferometric sensing [1]. In real-world fiber-optic systems, the importance of polarization preservation extends beyond simply ensuring the presence of a polarization state; it is the ability to maintain this polarization through propagation and in the presence of environmental disturbances such as temperature changes, mechanical stress, bending, and twist that defines a fiber's utility in practical applications. The polarization extinction ratio (PER) is the key metric used to quantify the effectiveness of polarization maintenance [2]. A high PER ensures that the polarization state is retained without significant contamination from orthogonal polarization modes. In applications requiring high precision, maintaining a high and stable PER over a long distance and under environmental disturbance is crucial, as even small deviations in polarization can lead to errors in interferometric detections and data acquisition [3,4].

Polarization-maintaining solid-core fibers (PM-SCF) have been the workhose for delivering stable polarization for decades [5,6]. Yet the very fact that light is confined in silica imposes intrinsic constraints on performance, including Kerr & Raman/Brillouin optical nonlinearity effect [7], laser-induced damage & mode instability [8], Rayleigh backscattering noise [9], thermal & mechanical birefringence drift [10,11], radiation-induced attenuation [12] and magnetic-optic Faraday effect [13]. Such issues limit PM-SCFs in applications like fiber-optic gyroscopes, high-power coherent laser systems, high-precision timing and frequency transfer, quantum information and communication technologies.

These fundamental limitations of PM-SCF have motivated the migration toward hollow-core fibers (HCFs), where light propagates predominantly in an air environment [14-16]. HCFs suppress optical nonlinearities by up to three orders of magnitude [17,18], raise damage thresholds beyond multi-GW cm$^{-2}$ levels [19], reduce Rayleigh backscattering by over 30 dB [20,21], and exhibit strong resistance to radiation and electromagnetic fields [22,23], opening new possibilities for high-power and high-fidelity photonic systems. Paradoxically, the very feature that gives HCFs their benefits, minimal light–matter interaction, makes it difficult to introduce birefringence for polarization maintenance [24]. Traditional methods of inducing birefringence, such as applying stress or introducing structural asymmetry, rely precisely on the fiber's sensitivity to external perturbations, which is fundamentally minimized in hollow-core designs. As a result, achieving polarization maintenance in HCFs remains an exceptionally challenging endeavor.

Early attempts at creating PM employed photonic bandgap (PBG) structured HCF [25-27]. Their performance was limited by several drawbacks: narrow operational bandwidth, high transmission loss, reliance on specific bending conditions for single-mode operation and complex structural designs. Recent years have witnessed remarkable advances in a more advanced HCF type: anti-resonant HCF (ARF) [15,16], achieving transmission losses below the Rayleigh scattering limit of conventional silica fiber [28,29]. Compared with PBG fibers, ARF offers much wider operational bandwidths and superior mode purity [16,29]. However, the pursuit of effective polarization preservation remains equally challenging.

As illustrated in Figure 1, two prevailing design strategies have been explored to achieve polarization preservation in ARFs, each entailing intrinsic trade-offs. (1) Non-PM ARFs. These fibers lack built-in slow/fast axes but can exhibit extremely high PER under ideal, static conditions and offer intrinsic immunity to thermal perturbations [30-32]. However, without a defined birefringence to pin the polarization, they are highly sensitive to mechanical disturbances. Even modest bending or a slight reorientation of the coil can randomize the output polarization state, and the principal polarization axes can rotate with wavelength [33-34]. (2) Birefringent polarization-maintaining ARFs (PM-ARFs). To stabilize polarization against perturbations, these fibers incorporate structural asymmetries to create birefringence. Since the hollow core itself doesn't naturally support birefringence, it is typically generated by leveraging the surrounding silica membranes [24,35]. This is achieved by introducing asymmetrically distributed cladding membranes and reducing the core size, thereby enhancing the modal overlap between them and yielding measurable refractive index differences between orthogonal polarizations [36-39]. Our previous work showed that a birefringence on the order of $10^{-5}$ to $10^{-4}$ is sufficient to lock polarization to fixed spatial axes and impart resilience to bending and twisting over tens of meters of fiber [38,39].

While this approach appears to offer an ideal solution, a critical limitation arises in practical implementations that is not easily captured in short length of fiber or by simulations: distributed polarization crosstalk accumulated along the fiber [30,40]. Many factors, like structural non-uniformities, microbending effects, and external perturbation can give rise to cumulative mode coupling during propagation. Paradoxically, the very design choice intended to enhance birefringence, namely reducing the core size, also increases the sensitivity of guided modes to such imperfections. This cross-coupling is quantified by the polarization crosstalk coefficient $h$. Note polarization crosstalk also exists in PM-SCF but to a much lesser degree (A comparative analysis will be presented in the Discussion section). Our recent distributed measurements reveal spatially distributed polarization crosstalk in a 20 µm-diameter core PM-ARFs with $h$ as large as $10^{-5}$ m$^{-1}$ in straight fiber, rising above $10^{-3}$ m$^{-1}$ under bending and stress [40]. Such distributed coupling is accumulative: as fiber length increases, even small $h$ leads to significant cross-polarization power transfer, progressively degrading the PER. Consequently, state-of-the-art birefringent PM-ARFs has yet demonstrated PER above 30 dB over propagation distance beyond 100 m. The challenge of delivering practically high and stable PER, both in short and long distance, and with environmental robustness remains a major roadblock in the development of practical PM-HCFs.

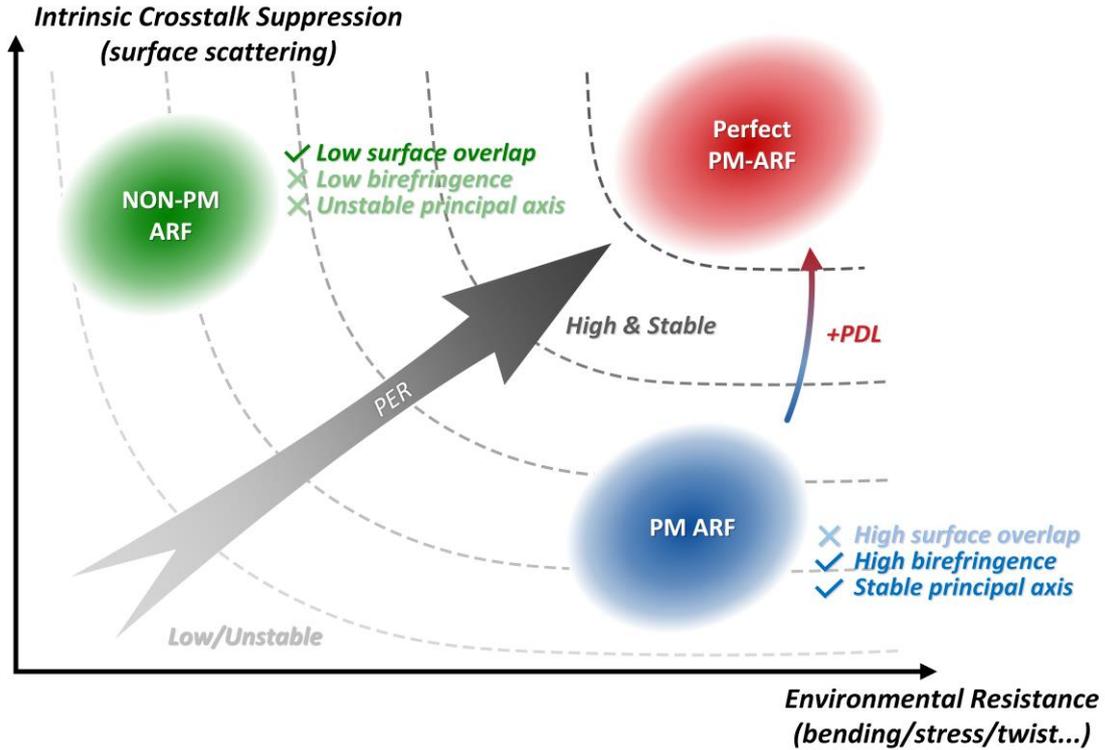

Fig. 1 Towards perfect PM-ARF: Conceptual comparison of different polarization preserving strategies for ARFs. Non-PM ARFs rely on large symmetric cores with low surface overlap to minimize polarization crosstalk but fail under bending or environmental perturbations (unstable). PM-ARFs with build-in birefringence use structural asymmetry and smaller cores to lock polarization axes, yet distributed polarization crosstalk accumulates with length because of its high surface overlap with silica cladding. The proposed PDL-enhanced BI-ARF design integrates engineered PDL with birefringence. This architecture suppresses accumulated crosstalk and enables high and stable PER over long-distance and under environmental perturbation—a pathway to the ideal "perfect PM-ARF."

In this work, we introduce a new design paradigm — polarization-differential loss (PDL) to break this impasse. By introducing and engineering proper PDL into a birefringent ARF, we passively suppress cross-coupling over long distances, allowing for the maintenance of a high and consistent PER throughout the fiber span. This is achieved by introducing phase-matching of one polarization mode with the leaky cladding resonance to substantially increase its loss, while maintaining the orthogonal polarization low-loss, yielding a large differential loss [41]. This engineered imbalance ensures that any optical power coupled into the unwanted polarization is rapidly attenuated along the fiber, effectively quenching polarization cross-talk. Using this PDL-enhanced approach, we demonstrate long-distance fibers with consistently high PER, despite substantial crosstalk coefficients ($h \sim 10^{-5}$ m$^{-1}$-$10^{-4}$ m$^{-1}$). In one of the fiber samples, we achieve a PER of ~30 dB across a ~45 nm bandwidth for a 725 m length with a minimum loss of 8.5 dB/km under a maximum PDL of 456 dB/km. In another shorter 102 m sample, we achieved an even higher PDL of ~1800 dB/km (at the expense of higher base loss) and realized ~40 dB PER over a ~50 nm bandwidth. These levels of polarization extinction are unprecedented in HCFs of such length, and importantly they remain stable under bending and other perturbations, addressing the key limitations of prior PM-HCF designs. This PDL-enhanced HCF architecture represents a practical and scalable pathway toward real-world deployment, paving the way for applications in high-precision gyroscopes [42], high-power coherent laser systems [43], high-precision timing and frequency transfer [44], and beyond.

## 2. Principle
### 2.1 PDL as an intrinsic PER-stabilization mechanism

A quantitative framework is first established to evaluate how deliberately engineered PDL mitigates polarization crosstalk and enhances PER in PM-ARFs. Figure 2a illustrates the basic problem: along any birefringent fiber, residual coupling $h$ causes a small fraction of optical power to continually transfer into the orthogonal polarization, whose cumulative effect ultimately limits the achievable PER [6,45].

In the absence of intentional loss asymmetry, according to the coupled-power model (see Methods), the PER evolves as [6],

$$PER = -10 \cdot \log\left(\frac{P_x}{P_y}\right) = -10 \cdot \log\left(\tanh(h \cdot L)\right), \quad (1)$$

where $P_y$ and $P_x$ denote the powers of the excited mode and coupled mode respectively. This predicts the power of unwanted polarization component grows along the fiber length ($L$), ultimately driving the PER towards 0 dB as $L$ extends to infinite. In practical terms, with $h$ on the order of $10^{-5}$ m$^{-1}$ to $10^{-4}$ m$^{-1}$, maintaining PER > 20 dB over long range is unattainable with birefringence alone.

When intentional PDL is introduced, the accumulated part of the polarization crosstalk can be effectively attenuated through the fiber (see Fig. 2a), fundamentally changing the picture. Under practical conditions, the PER evolves as (see Methods),

$$PER = -10 \cdot \log\left(\frac{P_x}{P_y}\right) \approx -10 \cdot \log\left[\frac{10 \cdot h}{\alpha_{PDL} \cdot \ln 10} \cdot \left(1 - 10^{-\frac{\alpha_{PDL} \cdot L}{10}}\right)\right], \quad (2)$$

here $\alpha_{PDL} = \alpha_x - \alpha_y$, where $\alpha_x$ and $\alpha_y$ are the loss of $x$- and $y$-polarization mode in dB/km, respectively. This reveals that the attainable PER depends on the differential loss $\alpha_{PDL}$ instead of the absolute loss of the low-loss axis. This enables a design strategy that imposes modest loss on the working axis while assigning very high loss to the orthogonal polarization. Figure 2b illustrates two representative cases. With $h = 10^{-4}$ m$^{-1}$, maximum PER of approximately 20 dB, 30 dB and 37 dB can be achieved with $\alpha_{PDL}$ values of 50 dB/km, 500 dB/km and 2000 dB/km, once the fiber length exceeds roughly 350 m, 30 m, and 8 m, respectively. For a smaller coupling coefficient $h = 10^{-5}$ m$^{-1}$, the same PDL values yield even higher PER ceilings (~30 dB, ~40 dB, and ~47 dB, respectively). Even over short distance of ~10 m, the benefit of PDL is evident. Only for sub-meter devices such as in-line polarizers [46], ultra-high birefringence alone may suffice to achieve comparable PER. In the common case where PM-ARFs are employed as transmission media, incorporating sufficient PDL becomes a prerequisite for sustaining high PER over practical link lengths.

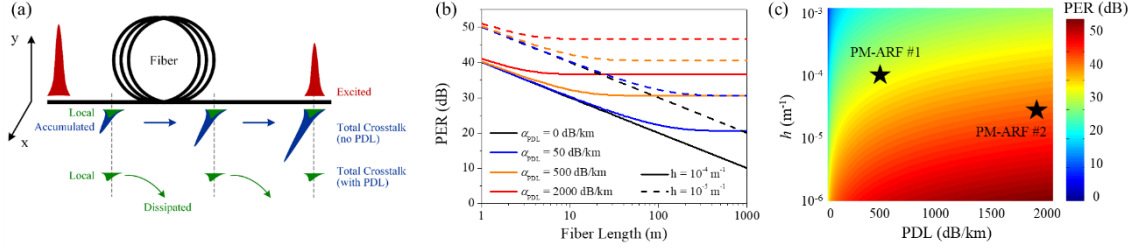

Fig. 2 The impact of PDL on PER. (a) Schematic view of polarization crosstalk accumulated in a fiber. (b) Estimated PER vs fiber length with different value of PDL when $h = 10^{-4}$ m$^{-1}$ (solid curve) or $10^{-5}$ m$^{-1}$ (dash curve). (c) Estimated PER vs PDL and $h$ when the fiber length is sufficiently long. The pentagrams in the figure mark the parameter sets achieved by the two fibers developed in this work.

Another way to view the impact of PDL is to consider the steady-state PER in a long fiber. When the fiber length greatly exceeds the polarization coupling length $L \gg 1/\alpha_{PDL}$, Eq. (2) simplifies to:

$$PER = -10 \cdot \log\left(\frac{P_x}{P_y}\right) \approx -10 \cdot \log\left(\frac{10 \cdot h}{\alpha_{PDL} \cdot \ln 10}\right). \tag{3}$$

In this regime, the PER no longer depends on $L$ at all, but instead approaches a length-independent ceiling determined by the ratio of differential loss to coupling ($\alpha_{PDL}/h$). This is a key advantage of using PDL: it actively suppresses the accumulated crosstalk by removing the coupled light rather than just trying to average it out. In contrast, simply increasing birefringence (lowering $h$) can delay crosstalk buildup but not completely prevent the gradual PER degradation as length increases. Figure 2c shows the theoretical PER limits for various combinations of $h$ and $\alpha_{PDL}$. In this work, we implemented two levels of PDL (marked by the stars in Fig. 2c) into two ARF designs, achieving PER performance that aligns well with theoretical predictions.

**2.2 Fiber design**

To achieve a large PDL, we intentionally phase-match one polarization mode to a leaky cladding resonance while keeping the orthogonal mode in a low-loss anti-resonant state. In tubular ARFs, two types of cladding resonances can be leveraged: airy modes that resides in the cladding air channels, and dielectric surface modes localized at the silica membrane [47]. We avoid using the air-channel resonances [35] because their effective indices are extremely sensitive to bending and other disturbances, which could disrupt the phase-matching. Instead, our design exploits a dielectric surface mode resonance in the silica membrane – a resonance with steep, material-dominated dispersion that is much more stable against mechanical perturbations. This approach ensures that the induced PDL will persist even under practical fiber bending or handling.

Multiple design objectives need to be simultaneously considered for PM-ARF: loss characteristics, birefringence, PDL, bend resistance, high order mode suppression, and fabrication feasibility. Our previous work on the four-fold truncated double nested anti-resonant nodeless fiber (4T-DNANF) demonstrated a viable platform that balances low loss and high higher order mode suppression ratio (HOMER) [29]. Building on this foundation, we incorporate the "bi-thickness" concept validated in our recent four-fold semi-tube ARF [38,39], yielding the structure shown in Fig. 3a. Here, the truncated tubes introduce intentional asymmetry between the polarization modes to create birefringence, which is further enhanced by the small core design, while the dual nested tubes serve primarily to further reduce optical loss.

In ARFs, the cladding tube wall thickness $t$ determines the resonant wavelength: $\lambda_m \approx 2t\sqrt{n_{glass}^2 - 1}/m$, $(m = 1, 2, 3, ...)$, which delineates the anti-resonant transmission band [48]. Crucially, for a given core polarization mode, the resonant condition is governed predominantly by the tube walls perpendicular to its field orientation. Consequently, when the four truncated tubes are engineered with distinct wall thickness ($t_1$ and $t_2$ in Fig. 3a), a deliberate offset between the resonant region (green box for $t_1$ and yellow for $t_2$) is formed in the dispersion curve of the two polarization modes, as shown in Fig. 3b. At the longer-wavelength edge of the dispersion curve (circled in red dot), the thinner wall ($t_1 < t_2$) causes the $LP_{01x}$ mode to reach its resonance condition first, resulting in strong coupling to the leaky dielectric mode, whereas the $LP_{01y}$ mode, still at the edge of its anti-resonant band is only weakly perturbed. The polarization experiencing higher attenuation is directly determined by which of the two thicknesses, $t_1$ or $t_2$, lies closer to its corresponding resonant wavelength $\lambda_m$ at the operating wavelength, thereby giving rise to the substantial loss essential for the engineered PDL.

Figure 3c focuses on the coupling region and highlights a second, purely geometric parameter: field symmetry. Since here the PDL mechanism primarily targets $x$-polarization component, we analyze the quasi-TM polarized modal family. Along the x-axis, the $LP_{01x}$ mode shares even parity with the $TM_{2k+1,j}$ dielectric modes, enabling anti-crossing and strong coupling. In contract, the orthogonal $LP_{01y}$ mode exhibits odd symmetry along the same axis, forbidding its coupling to $TM_{2k+1,j}$. Instead, it may only interact with the $TM_{2k,j}$ modes, which lies outside the relevant spectral region (grey region) and thus fail to facilitate energy transfer even if their dispersion curves intersect. This interplay between resonance proximity (set by $t_1$ vs $t_2$) and symmetry-allowed phase matching therefore selects a single polarization for rapid leakage, generating the robust PDL necessary for PER-stabilization scheme introduced in section 2.1.

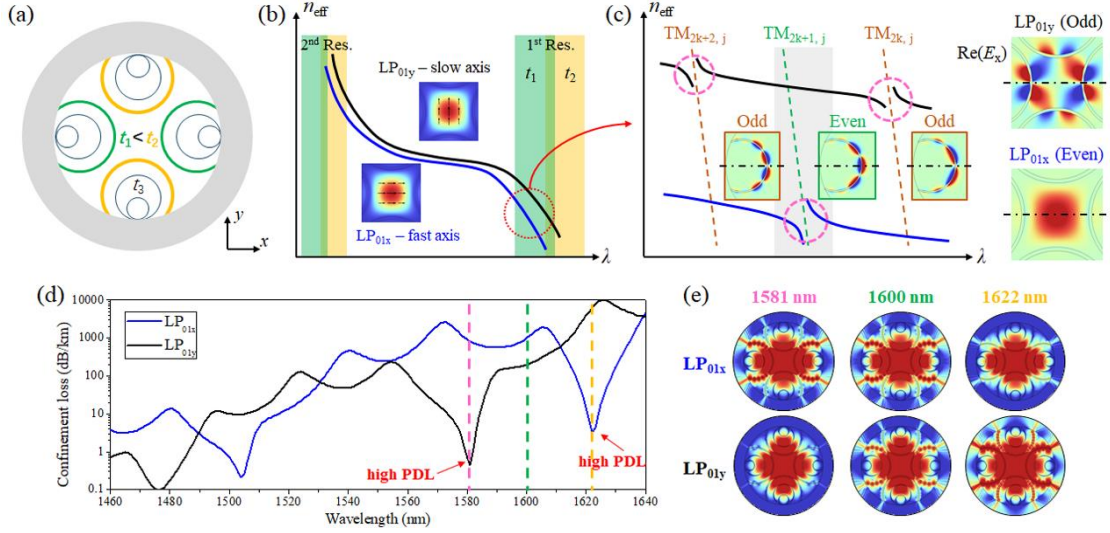

Fig. 3 Principle of obtaining high PDL in ARFs. (a) Schematic view of a 4T-DNANF with different wall thicknesses of cladding tubes. (b) Sketch of typical $n_{eff}$ of two polarization modes in an anti-resonant transmission window. Inset: the mode profile of $|E|$ and the polarization direction. (c) Enlarged display of the $n_{eff}$ of two polarization modes at longer-wavelength edge of the transmission window, when coupled with surface modes in the cladding tubes. Inset: the profile of the real part of $E_x$ for $LP_{01}$ core modes and quasi-TM surface modes. The red and blue indicates the phase of 0 and $\pi$, respectively. (d) The simulated loss spectra of two $LP_{01}$ modes of designed 4T-DNANF. (e) The simulated mode profiles of two $LP_{01}$ modes at 1581 nm, 1600 nm, and 1622 nm, respectively. Logarithmic scale is used.

To preliminarily validate our design, we conducted numerical simulations of the structure shown in Fig. 3a, featuring a core diameter of 19 μm, outer tube wall thicknesses of $t_1 = 0.95$ μm and $t_2 = 1.1$ μm, and inner tube wall thickness of $t_3 = 0.55$ μm. Finite element method (FEM) simulations were used to compute the loss spectra for both polarization modes near the long-wavelength edge of the second transmission window (Fig. 3d). The results reveal significant oscillations in the modal losses, with pronounced PDL observed at several specific wavelengths. Figure 3e displays the modal field distributions, revealing mode coupling characteristics at three representative wavelengths. At around 1581 nm, the $LP_{01x}$ mode strongly couples with the $TM_{17,2}$ mode in the thinner cladding tube ($t_1$), while the $LP_{01y}$ mode remains uncoupled due to symmetry protection, yielding a polarization loss ratio of 1000. In contrast, at around 1622 nm, the $LP_{01y}$ mode strongly couples with the $TM_{16,2}$ mode and exhibits high loss. At other wavelengths, such as 1600 nm, both polarization modes exhibit substantial losses as they each have corresponding coupling channels. Overall, the $LP_{11y}$ mode has smaller field overlap with the two thinner cladding tubes, making it more promising for achieving high PDL with low loss, such as the case at 1581 nm.

## 3. Results
### 3.1 PM-ARF#1: fabrication and characterization

Following the design concept, PM-ARF#1 was fabricated adopting a 4T-DNANF structure via the modified stack and draw method (methods). As shown in the scanning electron microscopy (SEM) image in Fig. 4a, the fabricated fiber features a core diameter of 19.4 µm. The four outer tubes exhibit membrane thicknesses of 1.06 µm and 0.96 µm along the fast axis, and 1.10 µm and 1.21 µm along the slow axis, respectively. These operate in the second AR band near 1550 nm. The nested tubes have membrane thicknesses ranging from 0.50 µm to 0.62 µm and operate in the first AR band. Only the two fundamental polarization modes are considered in this study, as the 4T-DNANF structure ensures sufficiently high loss of > 20 dB/m for higher-order modes (HOMs), as measured by spatially and spectrally resolved imaging [49,50] (see Supplementary Materials Section S4.1).

Transmission losses for both polarization modes were measured by aligning input polarizer to the x-/y-axes and applying the cut-back method from 860 m to 10 m (see methods and Supplementary Materials Section S3.1). As shown in Fig. 4b, the minimum loss of 1 dB/km occurs at 1460 nm, with negligible PDL across most of the transmission window. However, a substantial PDL emerges in the long-wavelength region from 1560 to 1610 nm. The black curve in Fig. 4c represents the PDL derived by subtracting the losses of the two polarization modes. At 1596 nm, the slow-axis polarization mode exhibits a minimum loss of 8.5 dB/km with a PDL of 280 dB/km and a polarization loss ratio of 33. At 1600.6 nm, the PDL reaches its peak value of 456 dB/km, while the slow-axis polarization mode exhibits a loss of 13.3 dB/km, corresponding to a polarization loss ratio of 35. The actual performance of the fabricated structure deviate from the simulated results of the ideal structure which predicts a PDL of 1000 dB/km and a loss of the low loss polarization mode of < 1 dB/km in Fig. 3d. This divergence likely stems from deviations in the fabricated fiber's structural thicknesses (see Supplementary Materials Section S1.2). In both axes, one of the two glass walls exceeds the designed specifications (0.95 µm and 1.1 µm respectively). Along the fast axis, the increased wall thickness reduces the filtering efficiency for the fast-axis polarization mode, consequently decreasing its attenuation. Conversely, along the slow axis, the thicker glass wall disrupts the anti-resonant condition for the slow-axis polarization mode, leading to increased loss. Optimization of the drawing process parameters could address these manufacturing tolerances in future iterations.

To ensure the accuracy of the cut-back measurements, two independent verification approaches were employed. First, FEM simulations were performed using the actual geometrical parameters obtained from SEM imaging. The simulated loss spectrum (dashed curves, Fig. 4b) shows excellent agreement with experimental measurements. Second, PDL was characterized on a 77 m fiber segment using an interferometric technique (see Methods and Supplementary Materials Section S3.4). In this approach, the incident light polarization was set to 45° to equally excite both polarization modes. Analysis of the output interference spectrum enabled determination of the intensity ratio between polarization modes through fringe visibility measurements. The derived PDL spectrum (blue curve, Fig. 4c) exhibits strong consistency with the cut-back results.

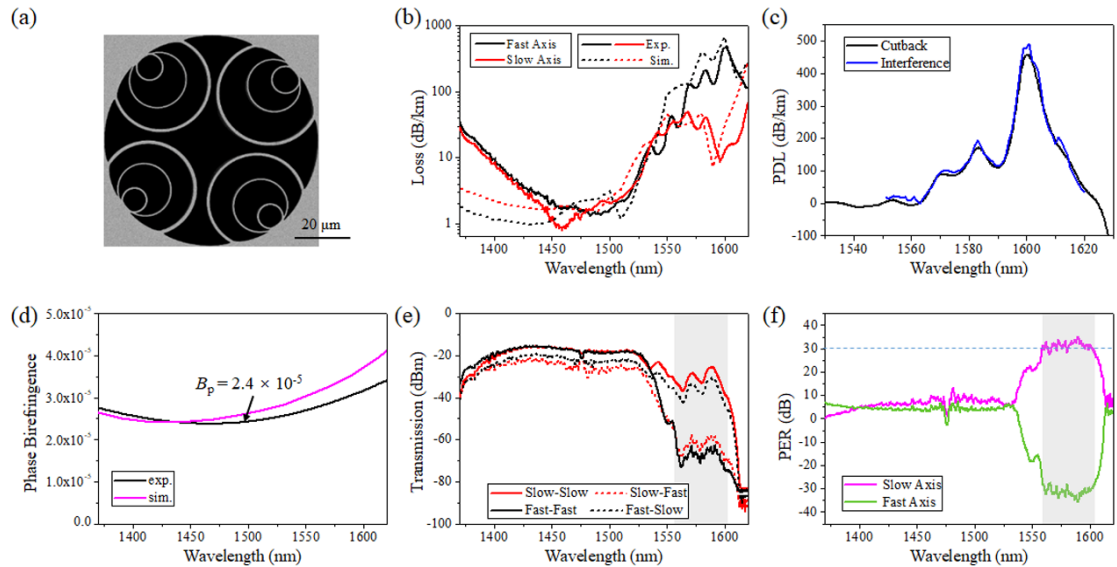

Fig. 4 Results of PM-ARF#1. (a) Cross sectional SEM picture. (b) Loss of two polarization modes measured by cut-back method (solid curves) and simulated based on the actual geometrical parameters obtained from SEM imaging (dashed curves), respectively. (c) Derived PDL by cut-back method and interferometric method, respectively. (d) Measured and simulated phase birefringence. (e) Crossed-polarizer transmission spectra under four excitation/detection configurations through the 725 m fiber. (f) The PER of two polarizations derived from (e).

Next we examine the polarization-related characteristics of this fiber. Phase birefringence measurements were conducted by analyzing polarization interference spectra at varying fiber lengths (Fig. 4d) showing values above $2.4\times10^{-5}$ throughout the whole transmission band, which agrees well with the simulation results. The critical performance metric, PER, was measured using a crossed-polarizer setup (see Methods and Supplementary Materials Section S3.2). Figure 4e presents four sets of transmission spectra measurements obtained from the 725 m long fiber under four excitation/detection configurations (combination of fast/slow axis excitation and fast/slow axis detection). The derived PER values (Fig. 4f) reveal distinct wavelength-dependent behavior. In the 1370 to 1520 nm wavelength range, corresponding to negligible PDL, the PER drops below 10 dB after 725 m of fiber propagation, indicating significantly degraded polarization-maintaining capability. The polarization crosstalk coefficient $h$ is estimated to exceed $10^{-4}$ m$^{-1}$. By contrast, in the high-PDL region spanning from 1560 to 1605 nm, a PER of approximately 30 dB across a 45 nm bandwidth over 700 m fiber length is achieved, demonstrating exceptional polarization preservation. Calculations in Fig. 2b, based on $h = 10^{-4}$ m$^{-1}$ and PDL = 500 dB/km, yield comparable PER values (~30 dB), indicating that the fiber has reached its PER ceiling-performance that can, in principle, be sustained over even longer lengths extending to kilometers. Notably, the fast-axis excitation reveals distinctive transmission behavior. Due to substantial propagation loss, the fast-axis polarized light is heavily attenuated over the 725 m span. Meanwhile, power coupled into the slow-axis mode, which experiences significantly lower loss, persists and ultimately dominates the output polarization state, which explains the negative PER values observed in Fig. 4f (see details in Supplementary Materials Section S2). The tradeoff for such exceptional PER performance is a marginal increase in propagation loss for the slow-axis polarization mode, rising from 1 dB/km in the non-PDL region to 8.5 dB/km in the high-PDL region.

## 3.2 Distributed polarization crosstalk

To further investigate polarization crosstalk in this fiber, we performed optical frequency-domain polarization crosstalk measurements on the 725 m long fiber (see Methods and Supplementary Materials Section S3.5) [40]. In these measurements, the linear excitation and detection polarization are fixed to be the low loss slow-axis (90°) and at 135°, respectively (Fig. 5a).

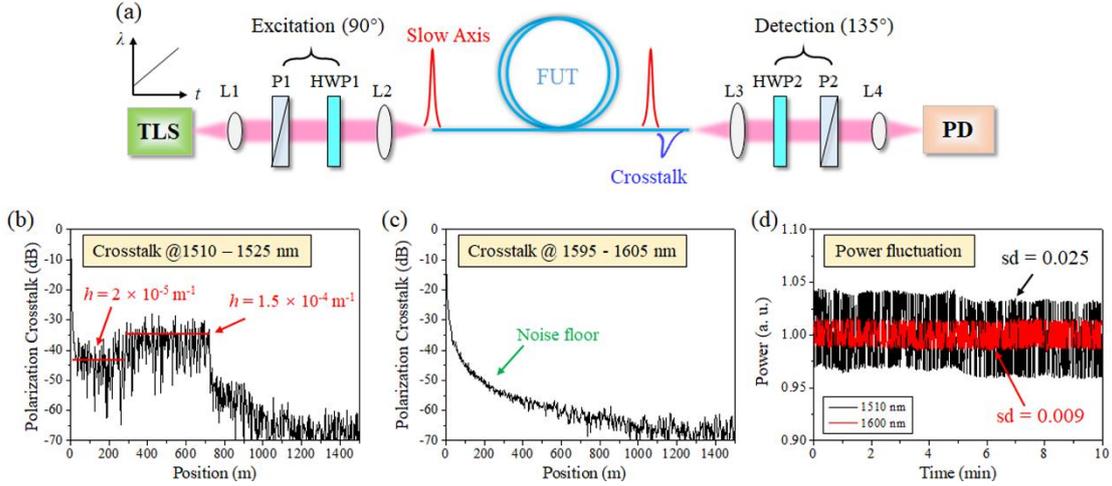

Fig. 5 Polarization crosstalk in PM-ARF#1. (a) Measurement setup. L1-L4: lens; P1/P2: polarizer; HWP1/HWP2: half-wave plate; TLS: tunable laser source; PD: photodetector; FUT: fiber under test. (b,c) Distributed polarization crosstalk at no-PDL and high-PDL wavelength ranges, respectively. (d) Transmission power fluctuations at no-PDL and high-PDL wavelength.

We first conducted distributed polarization crosstalk measurements in the non-PDL band, with Fig. 5b revealing severe polarization crosstalk within the 1510-1525 nm wavelength range. Over the 0-725 m fiber length, the polarization crosstalk exhibits distinct step heights between the first ~300 m segment and the subsequent ~400 m, clearly indicating a non-uniform crosstalk distribution. This may be attributed to internal structural non-uniformities or microbending effects along the fiber, though SEM imaging does not reveal noticeable microstructural differences between the two-fiber end-faces. Given the spatial resolution of 2.7 m in our distributed measurements, the $h$ value for these two fiber sections is estimated to be $2\times10^{-5}$ m$^{-1}$ and $1.5\times10^{-4}$ m$^{-1}$, respectively. These values are consistent with previous reported data for PM-ARFs [40], revealing the significant crosstalk issue inherent in PM-ARF designs. When substituting the average $h \approx 1\times10^{-4}$ m$^{-1}$ of this fiber segment into Eq. 2, the calculated output PER of ~11 dB aligns reasonably well with the experimentally measured ~8 dB (Fig. 3f). Next, we shifted the measurement wavelength range to 1595-1605 nm, with the results shown in Fig. 5c. Comparison with the results in Fig. 5b, the introduction of substantial PDL effectively suppresses the polarization crosstalk below the noise floor.

Another interesting phenomenon was the continuous temporal fluctuations of the transmission power observed during these measurements. With the setup maintained as in Fig. 5a and the output wavelength fixed, the detected optical power represents the interference between the slow-axis input light and the fast-axis crosstalk light. These power fluctuations were recorded over a 10 minutes period at both 1510 nm and 1600 nm, with the normalized intensity variations presented in Fig. 5d. The observed rapid fluctuations in polarization interference intensity are likely due to temperature-dependent phase shifts of the crosstalk light, as well as contributions from surface scattering and microbending. At PDL-enhanced region of 1600 nm, the standard deviation (sd) of these fluctuations is significantly reduced to 0.009, compared to 0.025 at the non-PDL 1510 nm wavelength, clearly demonstrating the suppressive effect of PDL on polarization crosstalk dynamics.

### 3.3 Resistance to bending

To evaluate whether the observed exceptional polarization preservation is robust against environmental perturbations—particularly mechanical disturbances—we conducted bending tests using a 77 m section of PM-ARF#1. At large bending radii, no discernible changes in the transmission spectra were observed, indicating negligible bending loss. As shown in Fig. 6a, measurable loss only emerges when the bending radius is reduced to 5 cm, with an additional loss remains below 0.1 dB/100 turns. In the high-PDL wavelength band, the bending loss spectrum exhibits oscillatory behavior and even displays negative values at certain wavelengths. This phenomenon is likely attributed to bending-induced modifications in the coupling between the polarization mode and the silica wall modes. Nonetheless, the fiber exhibits remarkably low overall bending loss, owing to its small core size.

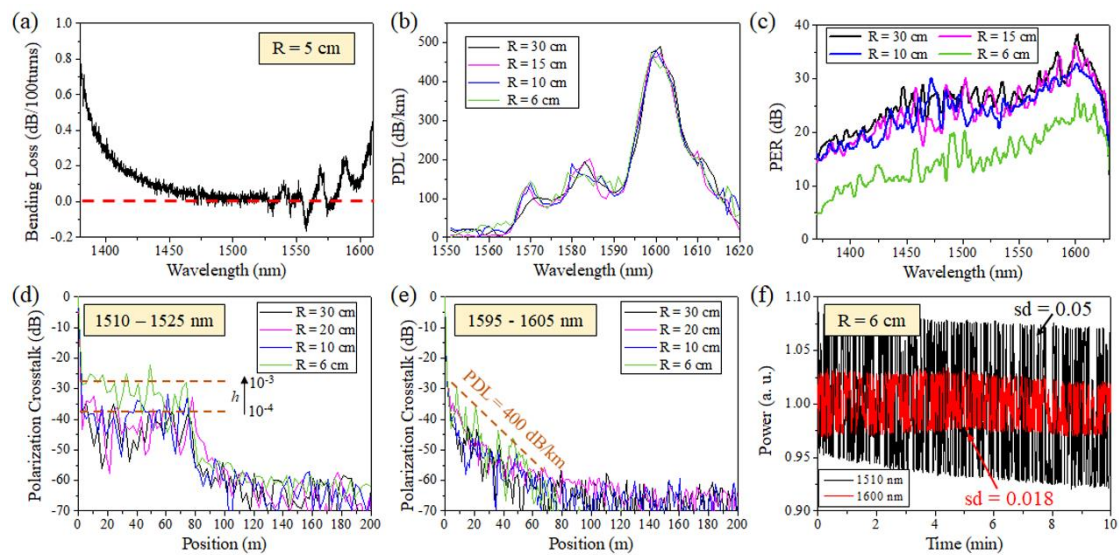

Fig. 6 Bending properties of PM-ARF#1. (a) Macro bending loss when the bending radius is 5 cm. (b) Measured PDL by interferometric method at different bending radii. (c) PERs of the low loss slow-axis polarization at different bending radii. (d-f) Distributed polarization crosstalk at (d) no-PDL and (e) high-PDL wavelength ranges at different bending radii, and (f) the corresponding transmission power fluctuations at certain wavelength.

We then evaluate the polarization-maintaining performance under bending. Figure 6c displays the PER spectrum under slow-axis excitation for this fiber segment. Given the relatively shorter fiber length of 77 m, the PER difference between the non-PDL and high-PDL bands is less pronounced. The results in Fig. 6c reveal no significant PER degradation at bending radii above 10 cm. When the radius is reduced to 6 cm, a noticeable drop in overall PER is observed, suggesting a substantial increase in polarization crosstalk. Notably, this bending tolerance is markedly superior to that of non-birefringent designs, where PER rapidly deteriorates under comparable mechanical perturbations.

To further identify the origins of crosstalk, distributed polarization crosstalk measurements were performed once again. As shown in Fig. 6d, in the 1510-1525 nm range, crosstalk remains distributed along the fiber for all bending radii. At a bending radius of R = 6 cm, the crosstalk coefficient increases dramatically from $10^{-4}$ $m^{-1}$ level to $10^{-3}$ $m^{-1}$ level, correlating well with the observed PER degradation. This increase is likely due to increased mode field overlap with the cladding tube walls under tight bending conditions, which substantially increase mode coupling. When shifting to the 1595-1605 nm wavelength range (Fig. 6e), polarization crosstalk remains largely suppressed across most bending radii. Only at R = 6 cm does the crosstalk slightly exceed the noise floor. The slope of crosstalk versus length under this condition yields an average PDL of ~400 dB/km, in close agreement with the value measured from Fig. 6b. Additionally, the fluctuation of transmission interference intensity over a 10 minutes period at R = 6 cm (Fig. 6f) were greater than those observed in the almost-straight-fiber condition (Fig. 5d), further indicating severe polarization crosstalk under bending. This is probably due to the enhanced surface scattering resulting from increased mode-surface overlap.

### 3.4 PM-ARF#2

To further enhance polarization-maintaining capability through increased PDL, Another PM-ARF (PM-ARF#2) following similar design principles but use smaller core diameter and more distinct glass wall thickness is fabricated. As shown in the SEM image in Fig. 7a, the fabricated fiber features a core diameter of 15.3 μm. The four outer tubes have membrane thicknesses of 1.02 μm and 0.98 μm on the fast axis, 1.29 μm and 1.31 μm on the slow axis respectively, while the remaining inner tubes exhibit membrane thicknesses ranging between 0.47 μm and 0.6 μm.

The loss spectrum obtained after cutback from 102 m to 10 m for the low-loss slow-axis polarization mode is shown in the red curve in Fig. 7b. We then measured the PDL using interferometry method, and obtained the fast-axis loss spectrum in the long-wavelength region by adding it to the slow-axis data (black curve in Fig. 7b). Figure 7b shows the loss spectra for both polarization modes. At 1586 nm, the PDL reaches its peak value of ~1800 dB/km, while the slow-axis polarization mode exhibits a loss of 48 dB/km, corresponding to a polarization loss ratio of 38.6. At 1593 nm, the slow-axis polarization mode achieves its minimum loss of 22.5 dB/km, with a polarization loss ratio of 55.5. The elevated loss of the low loss polarization mode can be attributed to enhanced mode coupling, caused by the smaller core diameter and greater wall thickness variations. This probably reflects a fundamental tradeoff in the design: achieving higher PDL comes at the cost of increased loss in the low loss polarization mode. Consequently, this fiber design is better suited for sub-100-meter applications where superior polarization-maintaining performance is required.

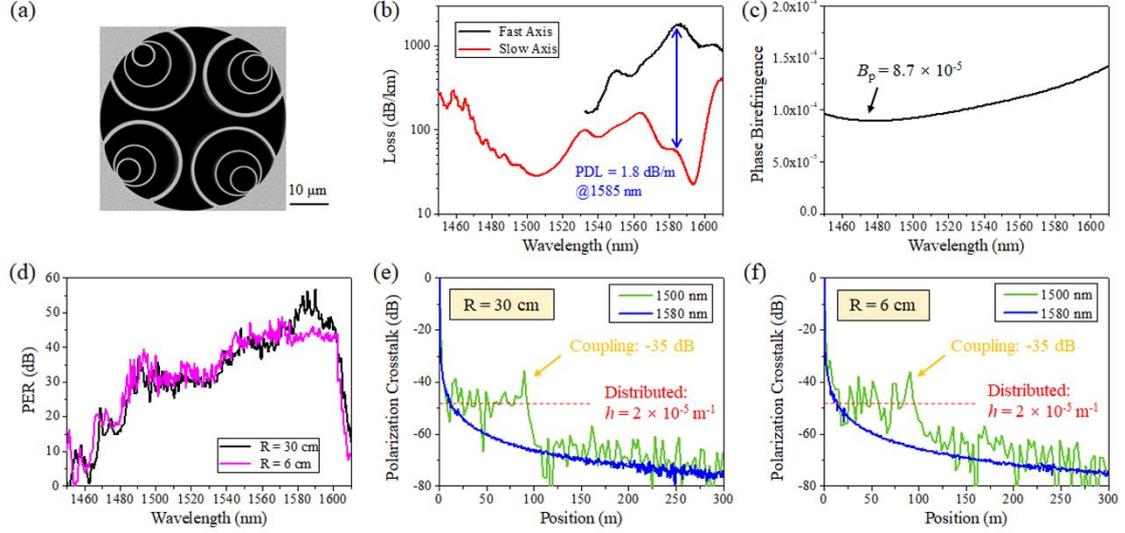

Fig. 7 Results of PM-ARF#2. (a) Cross sectional SEM picture. (b) Loss spectra of two polarization modes. (c) Measured phase birefringence. (d) PERs of the low loss slow-axis polarization at two bending radii. (e,f) Distributed polarization crosstalk at no-PDL and high-PDL wavelength ranges at two different bending radii, respectively.

Figure 7c shows that the fiber exhibits enhanced phase birefringence exceeding $8.7 \times 10^{-5}$. Thanks to the high PDL, the 102-meter fiber achieves a PER of ~40 dB across the 1550-1600 nm range (Fig. 7d), with a remarkable peak PER exceeding 50 dB within a 5 nm bandwidth around 1583 nm. Again, according to the PER stabilization mechanism, such performance is expected to be preserved over longer fiber length. In addition, as shown in Fig. 7d, such high PER of ~40 dB can be maintained when the fiber is coiled with a small bending radius of R = 6 cm. The almost same PER spectra at R = 30 cm and R = 6 cm indicates that the $h$ is maintained due to the higher phase birefringence, while the smaller PER at around 1583 nm maybe caused by a little degradation of PDL caused by bending. Distributed polarization crosstalk measurements in the no-PDL 1500-1515 nm range reveal an $h$ of $2 \times 10^{-5}$ m$^{-1}$ at both bending radii (Fig. 7e and 7f), which agree with the measured PER spectra. In contrast, at high-PDL wavelength range of 1580-1590 nm, the distributed polarization crosstalk is under the noise floor. Additionally, a notable ~ -35 dB polarization crosstalk is observed at the input coupling interface, likely due to the ellipticity of mode profile. Such crosstalk can also be effectively suppressed by the PDL mechanism.

## 4. Discussion and Conclusion

The approach of harnessing PDL fundamentally differs from that of PM-SCF which relies almost entirely on high built-in birefringence to preserve polarization. To illuminate the difference, it is useful to analyze the fundamental limits of polarization crosstalk suppression in different strategies.

The per-unit-length polarization coupling coefficient $h$, which critically governs polarization purity in optical fibers, can be expressed as [30]:

$$h = \left|\langle x|\Delta\varepsilon|y\rangle\right|^2 \left\langle \left|\Gamma(\beta_x - \beta_y)\right|^2 \right\rangle, \tag{4}$$

where $\left|\langle x|\Delta\varepsilon|y\rangle\right|^2$ quantifies the spatial overlap between the two polarization modes in presence of the permittivity perturbation tensor, $\Delta\varepsilon$. $\left\langle|\Gamma(\Omega)|^2\right\rangle$ represents the spatial power spectrum of local perturbation tensor with spatial frequency $\kappa_c$, which is inversely correlated with $(\beta_x - \beta_y)^2 + \kappa_c^2$. When the birefringence term $\beta_x - \beta_y$ exceeds $\kappa_c$, the birefringence can effectively suppress polarization crosstalk. This equation shows the origins of crosstalk can be divided into two categories: (1) intrinsic Rayleigh scattering, which contributes to the first term in Eq. (4), and (2) external perturbations (such as microbends, twists, and mechanical stress) that influence the second term via the perturbation spectrum. A classical coupled-mode theory result is that birefringence is effective at suppressing crosstalk from a given perturbation spectrum only if the birefringence wavenumber $\Delta\beta = \beta_x - \beta_y$ exceeds the spatial frequency content of that perturbation. In simple terms, a high $\Delta\beta$ (strong birefringence) averages out slow varying perturbations, but if perturbations vary faster along the fiber than $\Delta\beta$, they can still efficiently couple the modes. Low-frequency perturbations (large spatial period) are typically due to gentle external stresses, whereas high-frequency perturbations can arise from microscopic disorder like Rayleigh scattering centers.

In conventional SC-PMF, achieving very high birefringence is straightforward, and this readily suppresses most external disturbances. For example, $\Delta n$ of $3\times10^{-4}$ corresponds to a $\Delta\beta$ on the order of 1.2 mm$^{-1}$ at 1550 nm, enough to overcome typical perturbations (typically < 1 mm$^{-1}$ for microbending) [6]. As a result, well-made solid-core PM fibers can maintain high PER against environmental perturbations. However, Rayleigh scattering—intrinsic to the material and structurally unavoidable—sets a lower bound for $h$, usually around $3\times10^{-8}$ m$^{-1}$ [45]. According to Eq. 1, this intrinsic floor results in PER values on the order of 45 dB for a 1 km fiber with no added PDL. While this is sufficient for many applications, it still fails to guarantee perfect polarization isolation, as seen in state-of-the-art PANDA or elliptical-core fibers.

In contrast, achieving comparably low $h$ values in PM-ARFs is far more challenging. The intrinsic Rayleigh scattering originates from surface roughness at the silica-air interface. Although the typical root-mean-square (RMS) roughness is on the order of 0.2 nm [51], its contribution to scattering is significantly amplified by the high refractive index contrast between silica and air. Moreover, the perturbations due to surface roughness have very high spatial frequencies (> 2 mm$^{-1}$) [30,51]. No practical level of birefringence in a hollow-core design can directly suppress coupling from such high-$\kappa$ perturbations, because even the most birefringent ARFs reported with $\Delta n$ of $10^{-4}$ correspond to $\Delta\beta \lesssim 0.4$ mm. Even if we assume greater birefringence can be achieved, a fundamental trade-off is evident in Eq. (4): introducing birefringence in PM-ARFs inevitably requires strengthening the interaction of the core mode with the silica structure (e.g. smaller core, more asymmetric glass features). Those same design tend to increase the Rayleigh scattering contribution to $h$. In other words, a HCF cannot simultaneously have a large, low-scatter core and a small, high-birefringence core. Note here the non-BI approach mentioned in Fig. 1, which achieves $h$ value of $10^{-10}$ m$^{-1}$ by enlarging the symmetric core [30], is out of consideration because it fails to work in real-world handling.

This compromise was evident in the non-PDL operating regime of the two PM-ARF prototypes we developed. ARF#1, characterized by relatively weak birefringence, exhibits a spatial frequency of 0.1 mm$^{-1}$. Notably, distributed measurements reveal different $h$ values in its first 300 m ($2\times10^{-5}$ m$^{-1}$) compared to the subsequent 400 m ($1.5\times10^{-4}$ m$^{-1}$). With no observable structural differences under SEM, this discrepancy likely arises from different levels of microbending, which exceed the suppression capability of the current birefringence design. ARF#2 was originally engineered to emulate the conventional PM-SCF approach by employing a smaller core and introducing greater structural asymmetry, with the aim of increasing birefringence and reducing $h$. However, the measurements still show $h$ values of $2\times10^{-5}$ m$^{-1}$, due to the increased surface scattering. As a result, our measurements indicate that such birefringence-only designs fail to achieve high PER over long transmission distances.

The introduction of engineered PDL provides a fundamentally new route to overcome this trade-off. By selectively attenuating the unwanted polarization mode, PDL suppresses polarization coupling regardless of its origin (intrinsic or extrinsic). Even if $h$ remains non-negligible, any power that couples into the higher-loss polarization is quickly removed from the core, preventing it from propagating further and causing misalignment of polarization at the output. In effect, the fiber behaves like a single-polarization waveguide over long distances, without requiring extreme birefringence or perfectly uniform structure. Our experimental results demonstrate that a PDL-enhanced HCF can maintain PER levels (~30–40 dB) over hundreds of meters that far exceed those of prior HCF implementations, even under bent or stressed conditions. This advance has immediate implications for a range of polarization-sensitive systems. Fiber-optic gyroscopes, for instance, demand long coiled fibers with minimal polarization drift to avoid bias errors; the PDL-HCF approach can dramatically improve gyro stability by eliminating crosstalk-induced drifts. Similarly, coherent beam combining and interferometric sensor arrays benefit from maintaining a single polarization state over long fiber runs, as it prevents phase errors due to polarization rotation. High-precision timing and frequency transfer links, which currently are limited by polarization mode noise over long distances, could also see orders-of-magnitude improvement using PDL-stabilized HCFs. By breaking the previous constraints on HCF polarization performance, our work enables practical, long-distance delivery of polarized light with an unprecedented level of purity, opening the door for next-generation fiber systems in scientific and industrial applications.

**Methods**

**PER estimation**

When loss and polarization crosstalk exist, the power evolution of the two polarization modes along the fiber length can be described by the coupled power equation [36],

$$\begin{cases} \dfrac{dP_y}{dz} = -\dfrac{\ln 10}{10} \cdot \alpha_y \cdot P_y - h \cdot (P_y - P_x) \\ \dfrac{dP_x}{dz} = -\dfrac{\ln 10}{10} \cdot \alpha_x \cdot P_x + h \cdot (P_y - P_x) \end{cases}, \quad (5)$$

where $P_y$ and $P_x$ denote the powers of the excited mode and coupled mode respectively. $\alpha_x$ and $\alpha_y$ are the loss of $x$- and $y$-polarization mode in dB/km, respectively, and $z$ represents the longitudinal coordinate along the fiber length.

In the absence of intentional loss asymmetry (i. e. $\alpha_x = \alpha_y$), the solution to the equation is Eq. (1). When an extra attenuation $\alpha_{PDL} = \alpha_x - \alpha_y$ is applied selectively to the $x$-polarization mode, it can be assumed that $P_y \gg P_x$ and $\alpha_y \gg h$. Then, the solution to the equation is Eq. (2).

**Finite element method simulation**

The effective index, vector field distribution, and confinement loss of a polarization mode can be simulated using commercial finite-element solvers, such as COMSOL Multiphysics, with an optimized mesh size and perfectly matched layer. The geometrical parameters in Fig. 3 is ideal symmetry. The geometrical parameters in Fig. 4b and Fig. 4d was extracted from the SEM image of the fiber with some adjustment within the range of uncertainties.

**Fiber fabrication**

Fabrication of this dedicated PM-ARF structure begins with precision laser-cutting the full-length capillaries at defined angles. The resulting half-tubes are then accurately aligned with the inner, mid-sized, and small auxiliary tubes to build the stacked preform. This stack is thermally consolidated and drawn into a preform, which is subsequently pulled into fiber. During the draw, independent gas-pressure control is applied to four regions—the core, the truncated outer tubes, the intermediate tubes, and the smallest inner tubes—to preserve geometry and meet the intended optical specifications.

**Loss measurement**

The loss spectra of polarization modes are measured by cut-back method (see detail Supplementary Materials Section S3.1). The output light of a supercontinuum (SC-5, YSL, 470 nm to 2400 nm,) source is coupled into the fiber by using a set of lens. The input and output polarization are controlled by two calcite polarizers and two achromatic half-wave plates (HWP), enabling the detection of certain polarization modes. The transmission spectra are measured by an optical spectral analyzer (OSA, AQ6370D, Yokogawa, 600 nm to 1700 nm). Multiple measurements of the fiber reveal little variation in the measured loss spectra. Considering the excessive loss in the fast-axis polarization mode of PM-ARF#2, we adopted an alternative approach: first measuring the slow-axis mode loss and PDL separately, then summing them to obtain the loss for this polarization mode.

Macrobending loss measurement is conducted by comparing the transmission spectrum of a bent fiber with that of a quasi-straight fiber with the loop radius of 45 cm. After ensuring the excitation of low loss polarization modes, the output of the fiber was directly connected to the OSA through a magnetic clamp bare fiber adaptor (see detail Supplementary Materials Section S3.3).

For PDL, the first method for measuring PDL involves directly subtracting the losses of the two polarization modes, corresponding to the black curve in Fig. 4c. The second approach for measuring PDL utilizes an interferometric method. After ensuring a 1:1 excitation ratio of the two polarization modes using a polarizer and an achromatic HWP, the interference spectrum between them is measured. By analyzing this spectrum, the power ratio of the two polarization modes after fiber transmission can be determined, allowing calculation of their loss difference. The measurement results correspond to the blue curve in Fig. 4c.

**Phase birefringence measurement**

The phase birefringence is measurement by a polarization modal interferometer, which consisted of a supercontinuum source, two calcite polarizers, two half-wave plates (HWP), and an OSA (see details in Supplementary Materials Section S3.2). When the input and output linear polarization states are set to be 45° and 135°, respectively, the well-resolved spectral fringes of interference of two polarization modes can be acquired. Small segments of fiber is cut off continuously in mm-scale increments and the interferograms are recorded as a function of the fiber length. When the maxima/minima of the interferogram shift by one period as the fiber shortened by the length of $L_B$, phase birefringence can be calculated as $B_P = \lambda/L_B$, with $\lambda$ being the corresponding wavelength.

**PER measurement**

The polarization properties of our PM-ARFs were measured by a crossed polarizer setup, which consisted of a supercontinuum source, two calcite polarizers, two achromatic HWPs, and an OSA (see Supplementary Materials Section S3.2). By setting the input and detection polarizations to the $x$ and $y$ polarization modes respectively, four sets of transmission spectra can be acquired, corresponding to the results in Fig. 4e. The PER spectrum can then be obtained by subtracting these spectra.

**Distributed polarization crosstalk measurement**

The distributed polarization crosstalk was measured by optical frequency domain polarimetry method [40] (see details in Supplementary Materials Section S3.5). The measurement involved exciting only the low loss slow-axis polarization at the input end while detecting linear polarization at 45°, utilizing a tunable laser source (TLS) and a photodetector (PD) to acquire interference spectrum between the incident and crosstalk light, followed by Fourier transform to determine both the locations and magnitudes of crosstalk events.

The power fluctuation results in Fig. 5d is measured by the same setup. When the output wavelength of the TLS is fixed, the optical power were continuously recorded over a 10 minutes period.

**Conflict of interest**

The authors declare no competing interests.

**Acknowledgement**

This work was supported by the Scientific Research Innovation Capability Support Project for Young Faculty (ZYGXQNJSKYCXNLZCXM-I18 to Y. W.), National Natural Science Foundation of China (Nos. 62222506 and U21A20506 to Y. W., No. 62505111 to Y.S.), the Basic and Applied Basic Research Foundation of Guangdong Province (No. 2025A1515011728 to S. G.), the Guangzhou Science and Technology Program (No. 2024A04J9899 to S. G.), and the Major Key Project of Pengcheng Laboratory (W. D.).

**Author Contribution**

Y.W. conceived the study. Y.W. and S.G. designed the fiber structure. S.G. fabricated the fiber. Y.S. designed and supervised the fiber optical characterizations. X.W, Z.L, and R.H conducted the optical characterizations. X.Q. and Y.S. performed numerical simulations with support from W.D. and Y.W. Y.S. performed the PER estimation with assistance from W.D. and Y.W. Y.S. and Y.W. drafted the manuscript.